\documentclass{article}

\usepackage[nonatbib, preprint]{neurips_2020}

\usepackage[utf8]{inputenc} 
\usepackage[T1]{fontenc}    
\usepackage{hyperref}       
\usepackage{url}            
\usepackage{booktabs}       
\usepackage{amsfonts}       
\usepackage{nicefrac}       
\usepackage{microtype}      

\usepackage{amsmath}

\usepackage{graphicx}  
\usepackage{subfigure}
\usepackage{caption}

\title{A Computational Theory of Learning Flexible Reward-Seeking Behavior with Place Cells}

\author{
	Yuanxiang~Gao \\
	University of Electronic Science and Technology of China \\
	\texttt{gaoyuanxiang123@alu.uestc.edu.cn} \\
}

\begin{document}
	
	\bibliographystyle{plain}
	
	\maketitle
	
	\begin{abstract}
	An important open question in computational neuroscience is how various spatially tuned neurons, such as place cells, are used to support the learning of reward-seeking behavior of an animal. 
	Existing computational models either lack biological plausibility or fall short of behavioral flexibility when environments change. 
	In this paper, we propose a computational theory that achieves behavioral flexibility with better biological plausibility. 
	We first train a mixture of Gaussian distributions to model the ensemble of firing fields of place cells. Then we propose a Hebbian-like rule to learn the synaptic strength matrix among place cells. This matrix is interpreted as the transition rate matrix of a continuous time Markov chain to generate the sequential replay of place cells. During replay, the synaptic strengths from place cells to medium spiny neurons (MSN) are learned by a temporal-difference like rule to store place-reward associations. After replay, the activation of MSN will ramp up when an animal approaches the rewarding place, so the animal can move along the direction where the MSN activation is increasing to find the rewarding place. We implement our theory into a high-fidelity virtual rat in the MuJoCo physics simulator. In a complex maze, the rat shows significantly better learning efficiency and behavioral flexibility than a rat that implements a neuroscience-inspired reinforcement learning algorithm, deep Q-network.
	\end{abstract}
	
	\section{Introduction}
During the past decades, neuroscientists have discovered a variety of spatially tuned neurons, including place cells, head-direction cells, border cells, grid cells, etc \cite{Behrens2018, Moser2008}. Among them a representative cell type, the place cell, activates when an animal is in a particular location of the environment, serving as the representation of an animal's position. An important open question is how the activation patterns of place cells are used by an animal to support the learning of reward-seeking behavior. 

Some computational models \cite{Foster2000, Gustafson2011, Fremaux2013, Banino2018} proposed that the activation patterns of place cells are the set of basis functions for representing the value function in reinforcement learning (RL). However, reward-seeking behavior learned by RL can not flexibly adapt to the change of the rewarding location or the spatial layout. Another computational theory \cite{Muller1996} proposed that place cells and synapses among them can be modeled as a graph. To navigate to a rewarding place, an animal can follow the shortest path found in the graph from an initial place to the rewarding place. However, this model assumes that brains can implement the Dijkstra's shortest path algorithm \cite{Cormen2009}. There are no evidences that brains can perform such complex path search computations. 

In this paper, we propose a computational theory that achieves behavioral flexibility with better biological plausibility. First of all, we use a mixture of Gaussian distributions to model the ensemble of firing fields of place cells. The optimal firing field centers and dispersions are found by maximizing the total responsiveness of place cells to spatial positions. Then we propose a Hebbian-like rule to learn the strengths of synapses among place cells. We find that the synaptic strength between a pair of place cells not only encodes the spatial correlation of their firing fields but also determines the speed of activation propagation from one cell to another cell. As a result, the synaptic strength matrix among place cells is interpreted as the transition rate matrix of a continuous time Markov chain (CTMC), so the sequential replay of place cells during immobility \cite{Olafsdottir2017, Stella2019} is modeled as state transitions in the CTMC.

During replay, the strengths of synapses from place cells (PC) to medium spiny neurons (MSN) are learned by a dopamine-modulated temporal-difference like rule to store place-reward associations. We find that the PC-MSN synaptic strength encodes the spatial proximity between the firing field of each place cell and the rewarding place. As a result, during active movement, the activation of the MSN population will ramp up when an animal approaches the rewarding place.
In addition, the rate of change of the MSN activation along a direction encodes the motivation of the animal to move along that direction. Accordingly, we design a behavior policy that chooses a direction to move with a probability proportional to the derivative of the MSN activation along that direction.

We implement our theory into a high-fidelity virtual rat in the MuJoCo physics simulator \cite{Merel2020, Todorov2012}. In a complex maze environment, we compare the virtual rat with a virtual rat that implements a neuroscience-inspired RL algorithm, deep Q-network (DQN) \cite{Mnih2015, Hassabis2017}. The rat with our theory learns reward-seeking behavior over 50$\times$ faster than the DQN-based rat. When the rewarding location changes, the DQN-based rat requires over 10000 trials for relearning to navigate to the new location. In contrast, the rat with our theory can flexibly navigate to different rewarding locations without relearning costs. When the maze layout changes, the rat with our theory flexibly adapts its behavior to achieving rewards under the new layout, over 100$\times$ faster than the DQN-based rat. 
	\section{Theory}

\subsection{Learning firing fields of place cells}
Let $f(x|c_k)$ denote the normalized firing rate of the $k$-th place cell for an animal located at the position $x$. $f(x|c_k)$ is modeled by a 2D isotropic Gaussian density function denoted as $\mathcal{N}(x|\mu_k, \sigma_k^2)$, where the mean vector $\mu_k$ is the firing field center and the standard deviation $\sigma_k$ characterizes the dispersity of the firing field \cite{Foster2000, Gustafson2011, Banino2018, Muller1996}. For a population of $K$ place cells, the average of their firing rates at position $x$ is denoted by $\bar{f}(x) = \frac{1}{K} \sum_{k=1}^{K} \mathcal{N}(x|\mu_k, \sigma_k^2)$. $\bar{f}(x)$ quantifies the responsiveness of the population of place cells to the position $x$. For a dataset of positions $X = \{x_1, x_2, ..., x_N\}$ visited by a virtual rat exploring an novel environment, the optimal firing field centers and dispersities are found by maximizing the log-sum of average firing rate at each position as follows,
\begin{equation}
	\max\limits_{\{\mu_k, \sigma_k^2\}} \sum\limits_{n=1}^{N} \log [ \frac{1}{K} \sum_{k=1}^{K} \mathcal{N}(x_n|\mu_k, \sigma_k^2)].
	\label{MFP}
\end{equation}
The objective in Eq.~(\ref{MFP}) maximizes the total responsiveness of place cells to visited positions in the environment. The logarithmic transformations ensure that the responsiveness of place cells is distributed across all visited positions rather than focusing only on several particular positions. 

Solving the optimization problem in Eq.~(\ref{MFP}) is equivalent to finding a mixture of Gaussians that fits the dataset $X$ by the maximum likelihood principle \cite{Bishop2006}. We use the expectation-maximization (EM) algorithm \cite{Dempster1977, Bishop2006} to iteratively solve the optimal firing field centers and dispersities from arbitrary initial $\mu_k$ and $\sigma_k^2$. The E step updates the contribution ratio of the $k$-th place cell to the total firing rate at $x_n$,
\begin{equation}
\gamma_{nk}^{(t+1)} = \mathcal{N}(x_n|\mu_k^{(t)}, (\sigma_k^2)^{(t)})/\sum_j \mathcal{N}(x_n|\mu_j^{(t)}, (\sigma_j^2)^{(t)}), \forall n, k.
\label{Estep}
\end{equation}
The M step updates firing field centers and dispersities as follows,
\begin{equation}
\mu_k^{(t+1)} = \frac{1}{N_k} \sum\limits_{n=1}^{N} \gamma_{nk}^{(t+1)} x_n, \forall k,
\end{equation}
and
\begin{equation}
(\sigma_k^2)^{(t+1)} = \frac{1}{2N_k} \sum\limits_{n=1}^{N} \gamma_{nk}^{(t+1)} (x_n - \mu_k^{(t+1)})^{\rm T} (x_n - \mu_k^{(t+1)}), \forall k,
\label{Mstep}
\end{equation}
where $N_k = \sum_{n=1}^{N} \gamma_{nk}^{(t+1)}$. After each iteration, the total average firing rate in Eq.~(\ref{MFP}) is guaranteed to increase until convergence \cite{Bishop2006}. 

\subsection{Learning synaptic strengths among place cells}
During random exploration, a pair of place cells with nearby firing fields will often fire in close temporal order. The pair of recurrent synapses between this pair of place cells will undergo symmetric long-term potentiation (LTP) \cite{Isaac2009, Mishra2016}, whereas synapses between place cells with faraway firing fields will undergo no change in strength. These experimental observations are modeled as follows.

The strength of the synapse from the $i$-th place cell to the $j$-th place cell is denoted by the $ij$-th element of the $K\times K$ synaptic strength matrix $W$. For a sequence of positions $X' = \{x_{N+1}, x_{N+2}, ..., x_{N+T}\}$ visited by a virtual rat after learning place cell firing fields, the synaptic strength matrix is symmetrically updated by the following Hebbian-like rule \cite{Muller1996, Haykin1998},
\begin{equation}
W^{(t+1)} = W^{(t)} + \alpha (f(x_{N+t})^{\rm T} f(x_{N+t}) - W^{(t)}),
\label{Hebb}
\end{equation}
where $f(x_{N+t}) = \{f(x_{N+t}|c_k), k = 1,...,K\}$ is the row vector of place cell firing rates at position $x_{N+t}$. $\alpha$ is the learning rate. $W^{(0)}$ is chosen as a zero matrix.

The learning rule in Eq.~(\ref{Hebb}) is a stochastic approximation process \cite{Nevelson1976} that solves the following equation as $t \to \infty$,
\begin{equation}
W^* = \mathbb{E}[f(x)^{\rm T} f(x)],
\end{equation}
where the expectation is taken with respect to possible positions. The above equation is written elementwisely as follows,
\begin{equation}
W_{ij}^* = \mathbb{E}[f(x|c_i) f(x|c_j)], \forall i, j.
\label{corr}
\end{equation}
As shown by Eq.~(\ref{corr}), the synaptic strength between a pair of place cells encodes the correlation of their firing fields, which is determined by the proximity of their firing fields. Thus, the synaptic strength between a pair of place cells also characterizes the proximity of their firing fields.

\subsection{The sequential replay of place cells}
During wakeful immobility or sleep periods, place cells are spontaneously and sequentially reactivated, similar to the pattern of activations during previous movement periods \cite{Olafsdottir2017}. This process is called hippocampal replay. Recent studies further show that place cells are activated following a random walk like pattern during replay \cite{Stella2019, Pfeiffer2013}. These experimental observations inspired us to model the replay as follows.

During replay,  the activation is started from a randomly chosen place cell $i$. According to the quantal release of neurotransmitters  \cite{Branco2009, Gabbiani2017, Maass1999}, each spike of place cell $i$ induces the firing of a connected place cell $j$ with a certain probability proportional to the synaptic strength $W_{ij}$. Thus, the time for the activation of place cell $i$ to fire place cell $j$ approximately follows an exponential distribution $T_{ij}$ with mean $1/W_{ij}$. During these exponentially distributed waiting times, the fastest fired place cell will inhibit the activations of other place cells through fast inhibitory interneurons following a winner-take-all like mechanism \cite{Almeida2009, Colgin2016}. As a result, only the activation of the winning place cell is maintained and the activation is now transferred from place cell $i$ to the winning place cell. Similarly, the activation of the winning place cell will be transferred to another place cell following the waiting and competition process. This process is formally a continuous time Markov chain (CTMC) \cite{Norris1997} with the transition rate matrix given by the synaptic strength matrix $W$.

According to properties of exponential distributions \cite{Norris1997}, the minimum of the set of waiting times $H_i = \min_j T_{ij}, \forall i$ is still exponentially distributed with mean $1/\sum_{j\neq i} W_{ij}$. In addition, the probability that the waiting time of the place cell $k$ achieves the minimum is $P_{ik} = W_{ik}/\sum_{j\neq i} W_{ij}$. Accordingly, an equivalent approach to generate activation sequences is as follows. The activation is started from a randomly chosen place cell $i$. After an exponentially distributed holding time $H_i$, the activation is transferred to another place cell $k$ with probability $P_{ik}$. Similarly, the activation of the place cell $k$ is stochastically transferred to another place cell after a holding time $H_k$.

\subsection{Learning PC-MSN synaptic strengths}

The major functional role of replay is to learn place-reward associations \cite{Lansink2009, Sosa2020, Sadowski2016, Pennartz2004, Gomperts2015}. Such associations are stored in the synaptic strengths from hippocampal place cells (PC) to medium spiny neurons (MSN) in the striatum  \cite{Meer2011, LeGates2018, Trouche2019}. During replay, the activation of a place cell leaves a chemical trace (e.g., Ca$^{2+}$ concentration increase) at the synapse from the place cell to the population of MSN \cite{Kasai2021, Yagishita2014, Abrams1988}. The Ca$^{2+}$ is gradually taken up by organelles hence the chemical trace decays exponentially at the synapse \cite{Yagishita2014, Abrams1988}. Before the Ca$^{2+}$ is fully absorbed, if the dopamine concentration increases from a base level at the synapse, the synapse is strengthened proportional to both the chemical trace and the increase of the dopamine concentration \cite{Kasai2021, Yagishita2014, Gerstners2018}. These experimental observations are modeled as follows.

Let $\Delta t$ denote the refractory period \cite{Gerstner2014} between successive spikes of a place cell. Let $V^{(n\Delta t)}_i$ denote the synaptic strength from the $i$-th place cell to the MSN population at the $n$-th period. We call the place cell with maximal responsiveness to a given goal location $x_g$ as the goal cell \cite{Gauthier2018}. The index of the goal cell is determined by $\arg\max_k f(x_g|c_k)$. The activation of the goal cell at the $n$-th period is denoted by $g^{(n\Delta t)}$, which equals 1 if the activated place cell at this period is the goal cell, otherwise it equals 0. Let $e^{(n\Delta t)}_i$ denote the chemical trace at the synapse from the $i$-th place cell to the MSN population at the $n$-th period. Let $d^{(n\Delta t)}$ denote the deviation of dopamine concentration upon a base level at PC-MSN synapses at the $n$-th period. During replay, the PC-MSN synaptic strengths are updated by the following temporal-difference (TD) like rule \cite{Sutton2018, Gerstners2018},
\begin{equation}
	\left\{ 
	\begin{array}{lc}
		V^{[(n+1)\Delta t]}_i = V^{(n\Delta t)}_i + \alpha e^{(n \Delta t)}_i d^{(n\Delta t)}, \forall i, \\
		\\
		e^{[(n+1)\Delta t]}_i = e^{(n\Delta t)}_i \gamma^{\Delta t} + I_{c[(n+1)\Delta t]i}, \forall i, \\
		\\
		d^{(n\Delta t)} = g^{(n\Delta t)} + \gamma^{\Delta t} V^{(n\Delta t)}_{c[(n+1)\Delta t]} - V^{(n\Delta t)}_{c(n\Delta t)},\\
	\end{array}
	\right.
	\label{neoHeb}
\end{equation}
where $c(n\Delta t)$ and $c[(n+1)\Delta t]$ denote the index of the place cell activated at time $n\Delta t$ and $(n+1)\Delta t$, respectively. $I_{c[(n+1)\Delta t]i}$ is the indicator function that equals 1 if $c[(n+1)\Delta t]$ is $i$ and 0 otherwise. 
$\alpha$ is the learning rate and $\gamma$ is a discount factor close to 1 (e.g., 0.99). In Eq.~(\ref{neoHeb}), the chemical trace at each synapse decays geometrically with rate $\gamma$. In addition, if the place cell activated at time $(n+1)\Delta t$ is the $i$-th place cell, the chemical trace at the $i$-th PC-MSN synapse increases by 1. In Eq.~(\ref{neoHeb}), the deviation of dopamine concentration characterizes the activation of VTA dopamine neurons, which is a summation of following three signals, the signal on the disinhibitory pathway from the goal cell \cite{Luo2011}, the signal on the disinhibitory direct pathway from the MSN population and the signal on the inhibitory indirect pathway from the MSN population \cite{Keiflin2015, Morita2012}.

As derived in Appendix \ref{apdx1}, the learning rule in Eq.~(\ref{neoHeb}) is equivalent to a stochastic approximation process \cite{Nevelson1976} that solves the following equation as $n\to \infty$, 
\begin{equation}
V_i^* = \mathbb{E}[\sum\limits_{n=0}^{\infty} \gamma^{n\Delta t}g^{(n\Delta t)}], \ \ c(0) = i,
\label{goalGrad}
\end{equation}
where the expectation $\mathbb{E}[\cdot]$ is taken with respect to all possible activation transitions among place cells during replay. According to Eq.~(\ref{goalGrad}), $V_i^*$ is the expectation of the total discounted activation of the goal cell during replay started from the $i$-th place cell. For a place cell with its firing field geodesically close to the firing field of the goal cell, replays started from this place cell are more probable to frequently activate the goal cell in a near future. Thus, the synaptic strength from this place cell to the MSN population is stronger and vice versa. Therefore, $V_i^*$ encodes the geodesic proximity between the firing field of the $i$-th place cell and the firing field of the goal cell. 

\subsection{The generation of reward-seeking behavior}
Let $V(x)$ denote the activation of the MSN population when the animal is located at $x$ during active movement. At the $i$-th place cell's firing field center $\mu_i$, where the activation of other place cells is typically weak, the activation of the MSN population depends only on the $i$-th PC-MSN synapse strength $V_{i}$, so we have $V(\mu_i) \approx V_{i}, i = 1, ..., K$. 
Since $V(x)$ is a continuous function \cite{Meer2011}, $V(x)$ can be estimated by linearly interpolating the activation at three nearby firing field centers $\mu_i$, $\mu_j$ and $\mu_k$ as follows,
\begin{equation}
V(x)  = a_1 V(\mu_{i})+ a_2 V(\mu_{j}) + a_3 V(\mu_{k}), \\
\label{valFun}
\end{equation}
where nonnegative coefficients $a_1 + a_2 + a_3 = 1$ and $\mu_i$, $\mu_j$, $\mu_k$ are determined by first finding the Delaunay triangulation \cite{Berg2008} of the set of firing field centers then choosing the three vertices of the triangle enclosing $x$ as $\mu_i$, $\mu_j$ and $\mu_k$, respectively. $a_1$ equals the area ratio between the triangle $\Delta x \mu_i \mu_j$ and the triangle $\Delta \mu_i \mu_j \mu_k$ \cite{Press2007, Buss2003}. $a_2$ equals the area ratio between $\Delta x \mu_j \mu_k$ and $\Delta \mu_i \mu_j \mu_k$. $a_3$ equals the area ratio between $\Delta x \mu_k \mu_i$ and $\Delta \mu_i \mu_j \mu_k$. 

Let $\partial_{\theta}V(x)$ denote the directional derivative of $V(x)$ along the direction $\theta$ (the angle with $x$-axis) at location $x$. $\partial_{\theta}V(x)$ characterizes the dopamine release upon a base level while moving along the $\theta$ direction at location $x$ \cite{Kim2020, Gershman2014}, which determines the motivation for moving along that direction \cite{Silva2018, Howe2016}. Normalizing the motivation along each direction by a softmax function defines a probability distribution with respect to $\theta$ as follows,
\begin{equation}
P_{x}(\theta) = \frac{{\rm exp}(\beta \partial_{\theta}V(x))}{\int_0^{2\pi} {\rm exp}(\beta \partial_{\theta}V(x)) \mathrm{d} \theta},
\end{equation}
where $\beta$ characterizes the greediness level of behavior. When $\beta$ varies from zero to infinity, the behavior changes from purely random exploration to dopamine-maximization behavior. Since it's difficult to sample the above continuous probability distribution, we use a discrete probability distribution by considering $M$ uniformly spaced directions $\{\theta_m, m = 1, ..., M\}$ as follows,
\begin{equation}
	P_{x}(\theta_m) = \frac{{\rm exp}(\beta \partial_{\theta_m}V(x))}{\sum\limits_{m=1}^{M} {\rm exp}(\beta \partial_{\theta_m}V(x))}.
	\label{discDist}
\end{equation}
Dopamine release is a gating signal to initiate a period of movement rather than continuously modulating ongoing movements \cite{Silva2018, Howe2016}. Accordingly, the locomotion of the virtual rat is divided into periods of movements. During each movement period, the virtual rat first samples the discrete distribution in Eq.~(\ref{discDist}) and then keeps moving along the chosen direction until the end of this movement period. 

    \section{Experiments}

\subsection{Experimental setup} 

\begin{figure*}[h]
	\centering
	\subfigure[\label{rodent}]{\includegraphics[width=1.2in, height=0.9in]{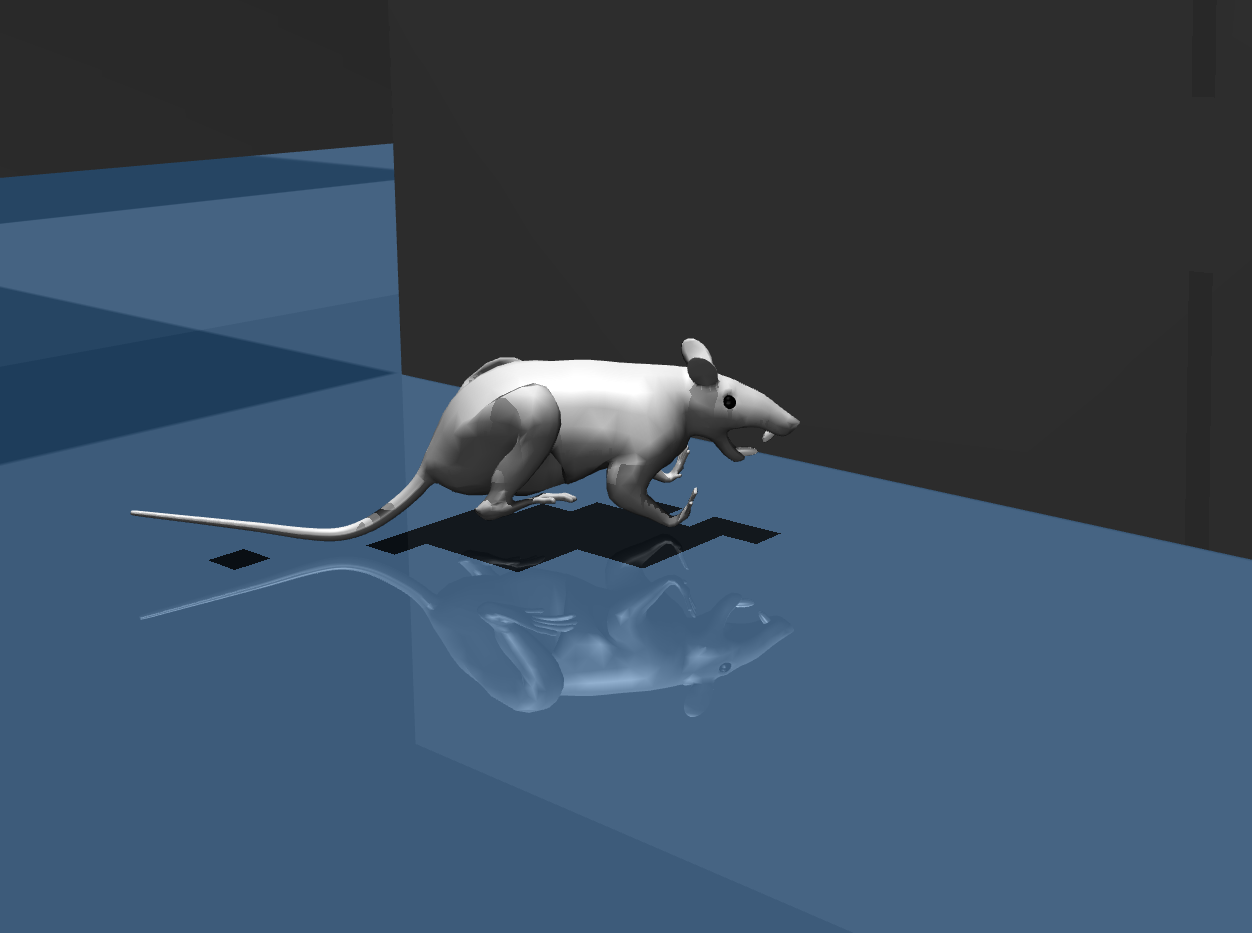}}
	\hfil
	\subfigure[\label{env1}]{\includegraphics[width=1.2in, height=0.9in]{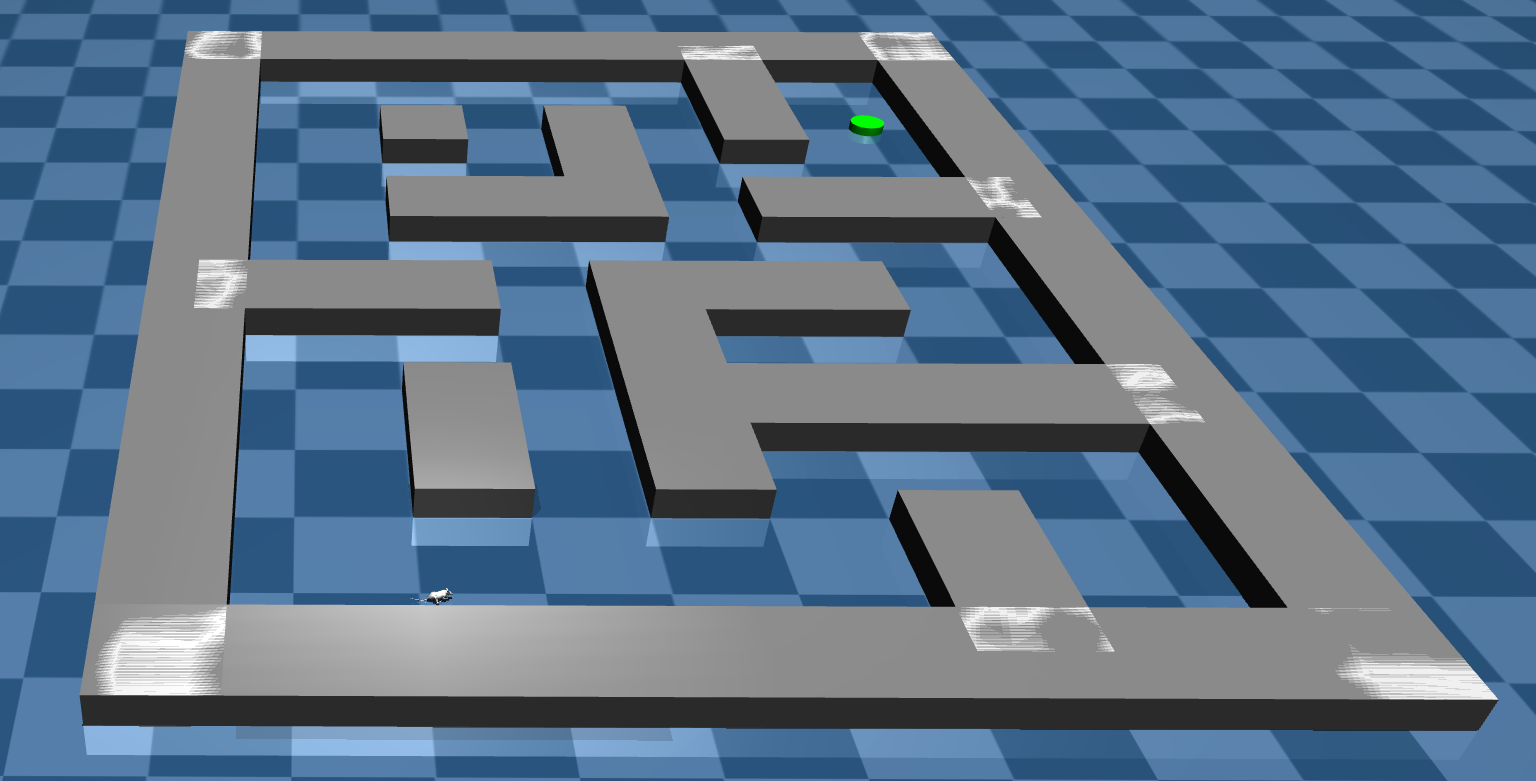}}
	\hfil
	\subfigure[\label{env2}]{\includegraphics[width=1.2in, height=0.9in]{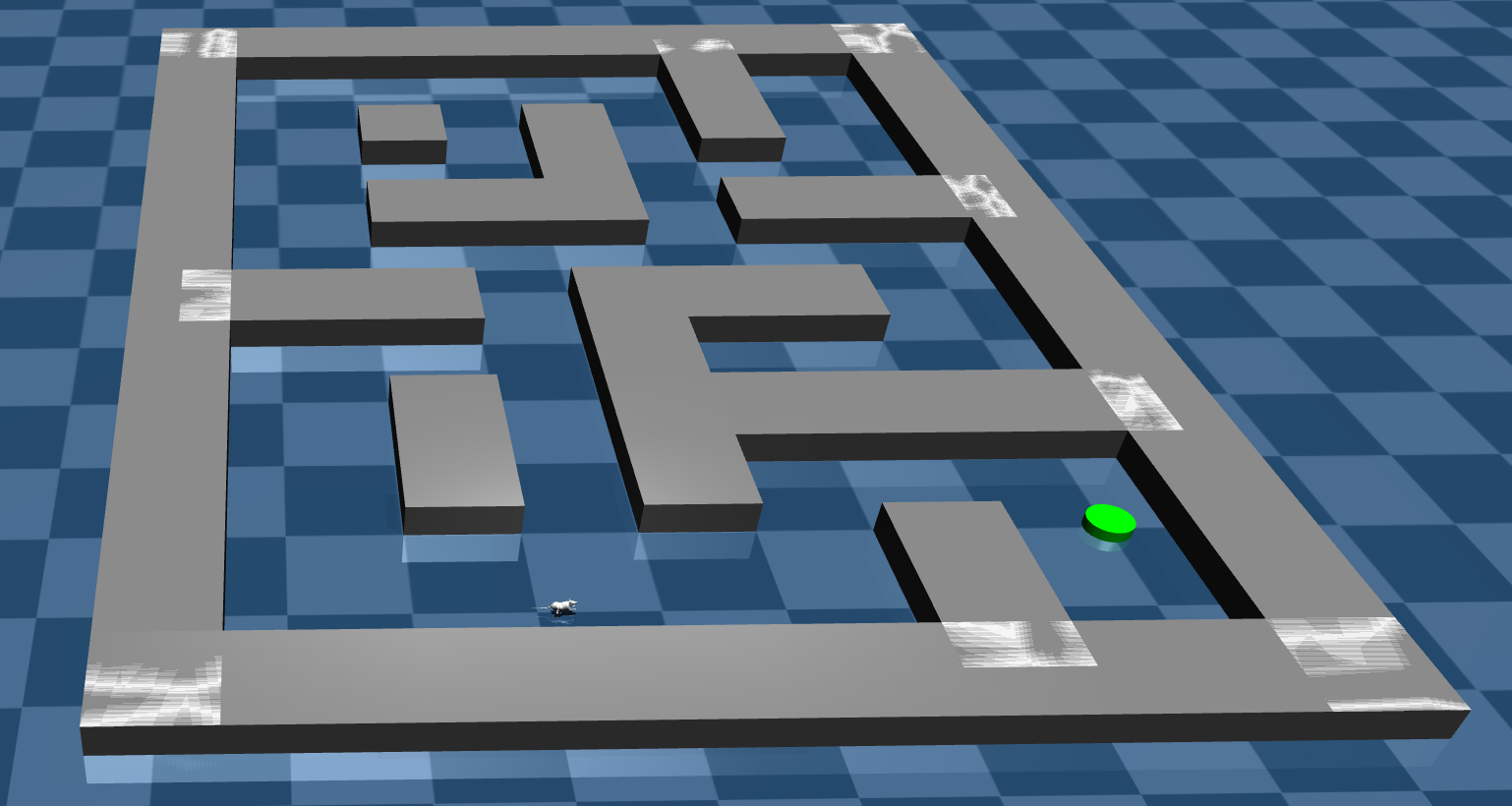}}
	\hfil
	\subfigure[\label{env3}]{\includegraphics[width=1.2in, height=0.9in]{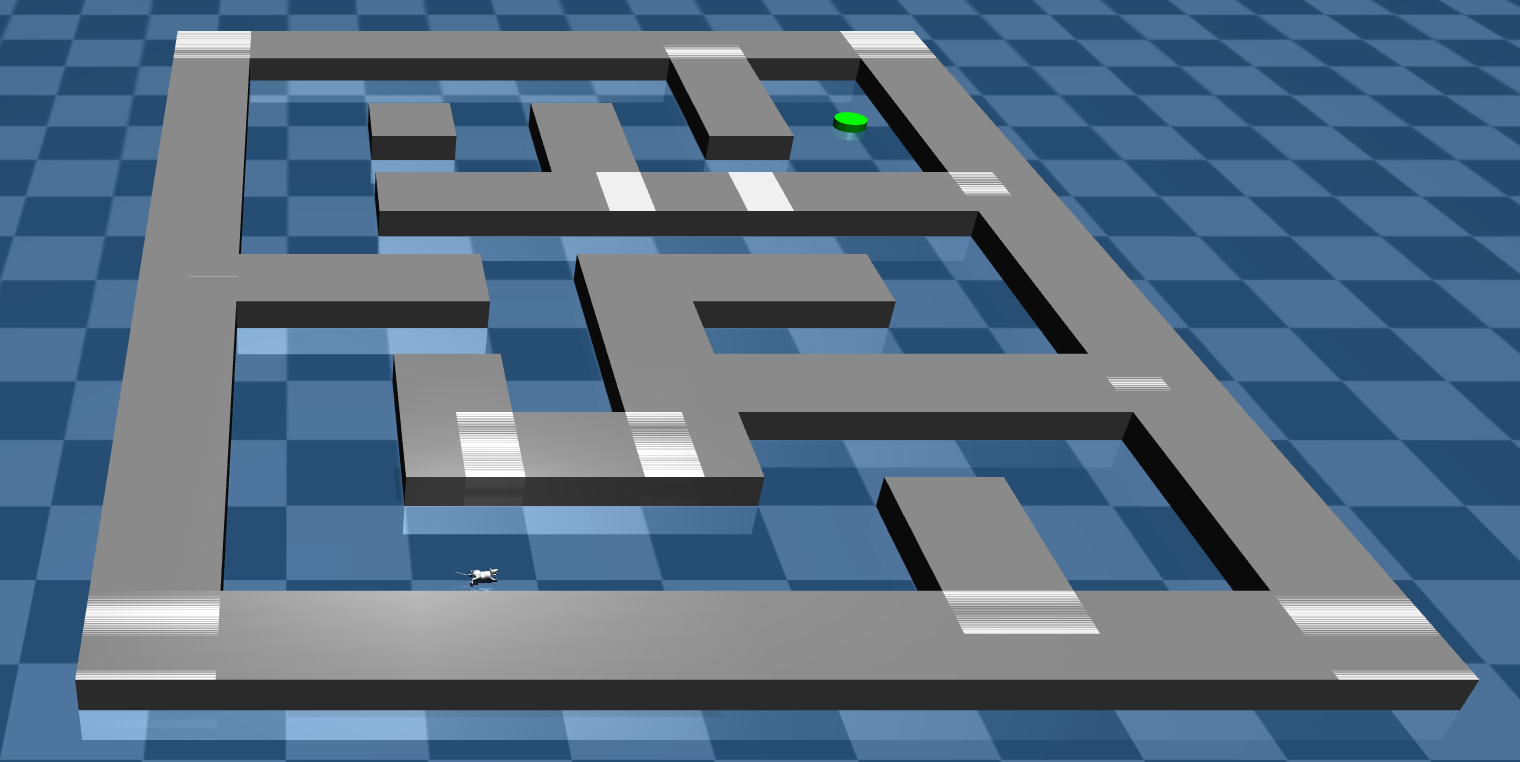}}
	\caption{The virtual rat and maze environments. (a) The virtual rat used in experiments.  (b) The maze environment used in the goal-fixed experiment. (c) The maze environment used in the goal-changing experiment. (d) The maze environment used in the maze-changing experiment.}
	\label{envs}
\end{figure*}

\textbf{Subject.} The subject used is a virtual rat that reproduces the anatomical features of seven Long-Evans rats in the MuJoCo physics simulator \cite{Merel2020, Todorov2012}, as shown in Figure \ref{rodent}. At each simulation step, the proprioceptive input to the virtual rat is an 148-dimension vector that includes the position, velocity and angular velocity of each joint. The motor output from the virtual rat is a 38-dimension vector that contains the torque applied to each joint. We use the TD3 deep RL algorithm \cite{Fujimoto2018} to pretrain the rat in an open field to master several basic locomotion policies, including running forward along the current body direction, turning the body by a certain angle then running forward. The trained turning angles are 45$^\circ$, 90 $^\circ$, 135$^\circ$, 180 $^\circ$, -45 $^\circ$, -90 $^\circ$ and -135 $^\circ$. Each basic locomotion policy is represented by a feedforward neural network that maps the proprioceptive input into the correct motor output to generate the desired locomotion. The pretraining details see Appendix \ref{pretrain}. For moving along a direction sampled from Eq.~(\ref{discDist}), the rat compares its current body direction with the sampled direction and invokes the closest locomotion policy to approximately run along the sampled direction for a period. 

For testing the proposed computational theory, we build a large scale 9$\times$11m maze in MuJoCo as shown in Figure \ref{env1}. We conduct following three experiments in the maze environment. 

\textbf{Goal-fixed experiment.} In the goal-fixed experiment, a location at the top right corner is set as the goal location, as shown in Figure \ref{env1}. The rat first randomly explores the maze for 100 epochs (trials) starting from random initial positions in the maze. Every epoch contains 4000 simulation steps. Before every 100 simulation steps, the rat randomly chooses a turning angle and runs along this direction in next 100 simulation steps. The positions visited during the first 50 epochs are used to learn the firing fields of 120 place cells by the EM iterations given in Eq.~(\ref{Estep})-Eq.~(\ref{Mstep}). After that, the positions visited during the last 50 epochs are used to learn the synaptic strength matrix of place cells by the Hebbian-like rule given in Eq.~(\ref{Hebb}). Then the rat reactivates place cells to learn the PC-MSN synaptic strengths by the TD-like rule given in Eq.~(\ref{neoHeb}). After that, the rat performs a series of test epochs with every epoch containing 4000 simulation steps. Before every 100 simulation steps, the rat samples a new direction from Eq.~(\ref{discDist}) and runs along this direction for next 100 simulation steps. Test epochs with the rat ending up within a 1$\times$1m square enclosing the goal location are treated as successful epochs. The success rate of the most recent 50 epochs is recorded as the performance measure of the rat.

\textbf{Goal-changing experiment.} In the goal-changing experiment, the exploration and learning phase is same as the goal-fixed experiment. However, after some test epochs, the goal location is changed from the top right corner to the bottom right corner as shown in Figure \ref{env2}. After goal changing, the rat reperforms replay to update the PC-MSN synaptic strengths with a new goal cell. Then the rat continues to perform test epochs. 

\textbf{Maze-changing experiment.} In the maze-changing experiment, the exploration and learning phase is also same as the goal-fixed experiment. However, after some test epochs, two passages in the maze are closed as shown in Figure \ref{env3}. After spatial layout changing, the rat randomly re-explores the maze for 100 epochs. The positions visited during the first 50 and last 50 epochs are used to update place cells' firing fields and synaptic strength matrix, respectively. After that, the rat reactivates place cells to update the PC-MSN synaptic strengths and then continues to perform test epochs.

We compare the rat equipped with our theory against rats that implement a neuroscience-inspired deep RL algorithm, deep Q-network (DQN) \cite{Mnih2015, Hassabis2017}. We design two variations of DQN algorithms, PC+DQN and GC+DQN.

\textbf{PC+DQN.} This baseline uses the experienced positions and head directions during the exploration phase to learn the firing fields of 120 place cells and 120 head direction cells. Then a series of training epochs is performed. Before every 100 simulation steps during each training epoch, the firing rates of place cells and head direction cells are fed into a feedforward neural network to predict the Q-values of the eight basic locomotion policies. During next 100 simulation steps, the policy with the maximal Q-value is used with probability $1-\epsilon$ or a randomly chosen policy is used with probability $\epsilon$. The reward is one if the rat is located within a 1$\times$1m square enclosing the goal location, otherwise the reward is zero. 
The Q-value network is trained by the DQN algorithm.

\textbf{GC+DQN.} This baseline uses firing rates of grid cells (GC) and head direction cells as input of the Q-value network.
The model used to compute firing rates of grid cells is provided in Appendix \ref{model-gc}. Except for the input, the GC+DQN baseline uses the same setup as the PC+DQN baseline.

Hyperparameters and their values used in experiments are summarized in Appendix \ref{hyperparas}.

\begin{figure*}[h]
	\centering
	\subfigure[\label{expers_exp1}]{\includegraphics[width=1.1in]{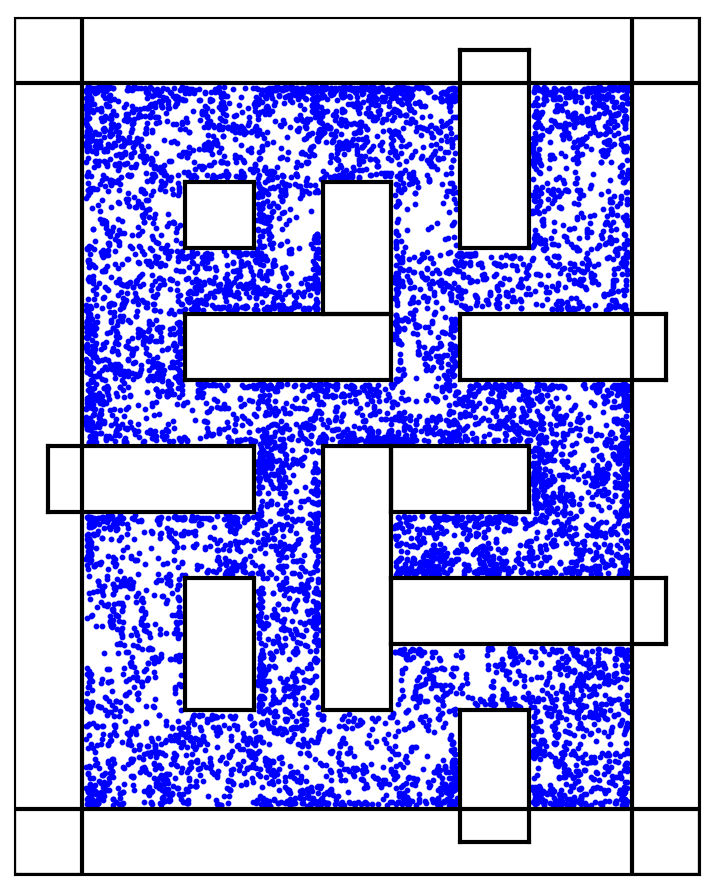}}
	\hfil
	\subfigure[\label{pcfield_exp1}]{\includegraphics[width=1.25in]{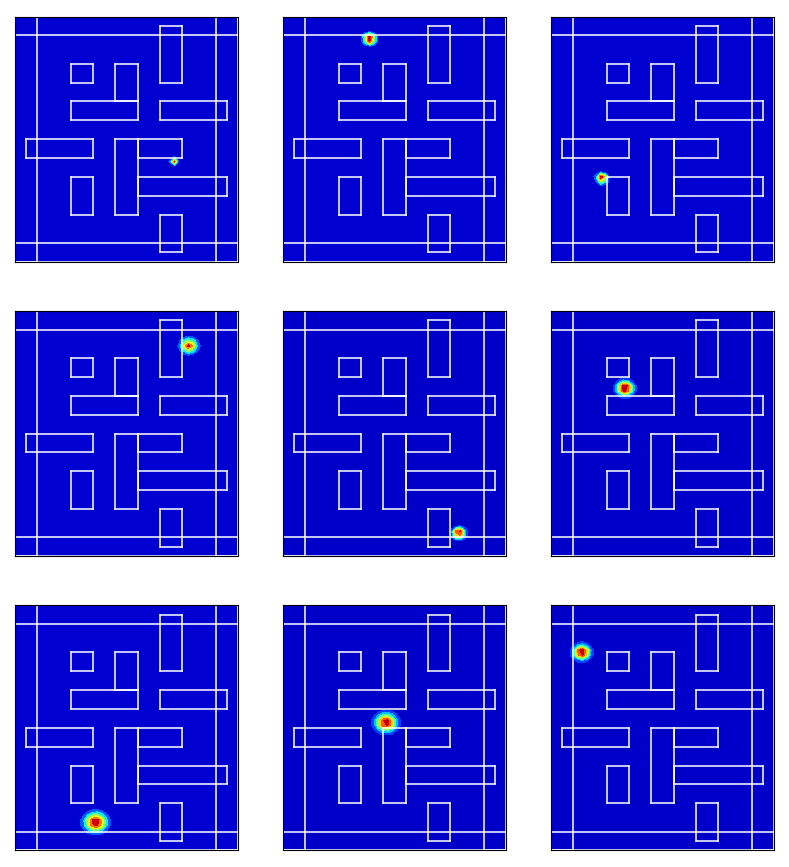}}
	\hfil
	\subfigure[\label{spatialcor}]{\includegraphics[width=1.25in]{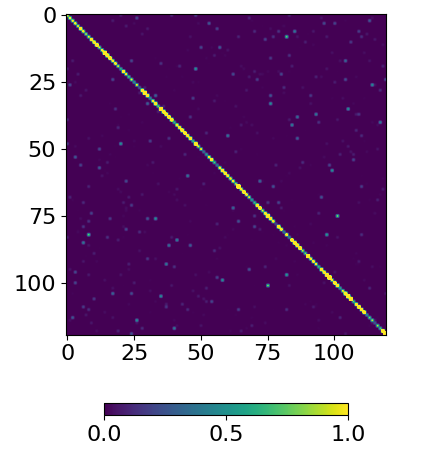}}
	\hfil
	\subfigure[\label{synstrenth}]{\includegraphics[width=1.2in]{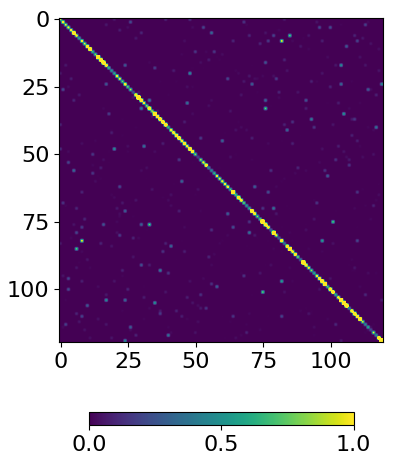}}
	\caption{Goal-fixed experiment. (a) Spatial positions visited during exploration. (b) Exemplary place cell firing fields. (c) The spatial correlation matrix of firing fields. (d) The synaptic strength matrix of place cells.}
	\label{experiment1a}
\end{figure*}

\subsection{Experimental results}

\textbf{Goal-fixed experiment.} Figure \ref{expers_exp1} shows the positions visited by the rat during exploration epochs. After the 100 exploration epochs, the maze is fully explored and the spatial experiences cover the whole maze. Figure \ref{pcfield_exp1} shows the firing fields of several place cells. Each place cell is selectively responsive to a restricted area of the maze. The firing fields of all place cells can cover the whole maze. Figure \ref{spatialcor} shows the correlation between each pair of firing fields computed by Eq.~(\ref{corr}). A firing field is typically highly correlated with several nearby firing fields. 
Figure \ref{synstrenth} shows the synaptic strength matrix of place cells learned by Hebbian-like learning. The synaptic strength matrix well characterizes the firing field correlations among place cells. 

\begin{figure*}[h]
	\centering
	\subfigure[\label{valMap_exp1}]{\includegraphics[width=1.4in]{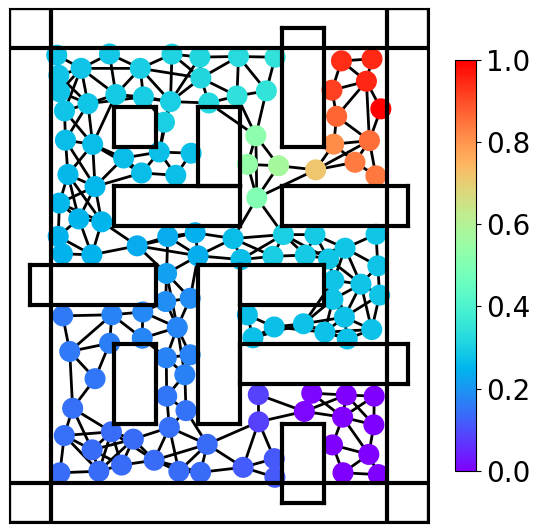}}
	\hfil
	\subfigure[\label{valFunc_exp1}]{\includegraphics[width=1.4in]{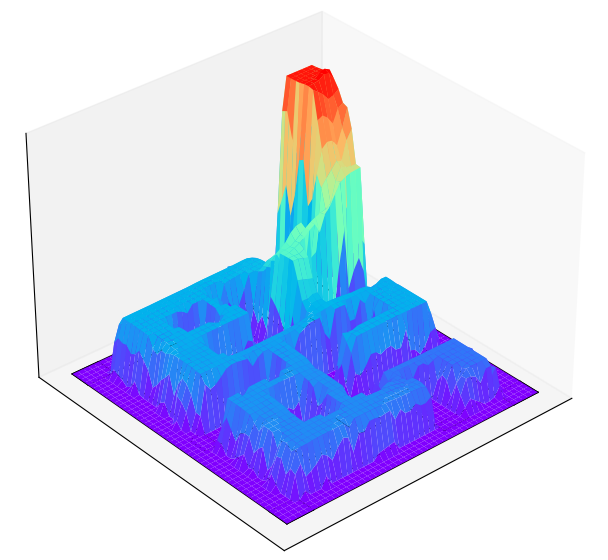}}
	\hfil
	\subfigure[\label{gradField_exp1}]{\includegraphics[width=1.15in]{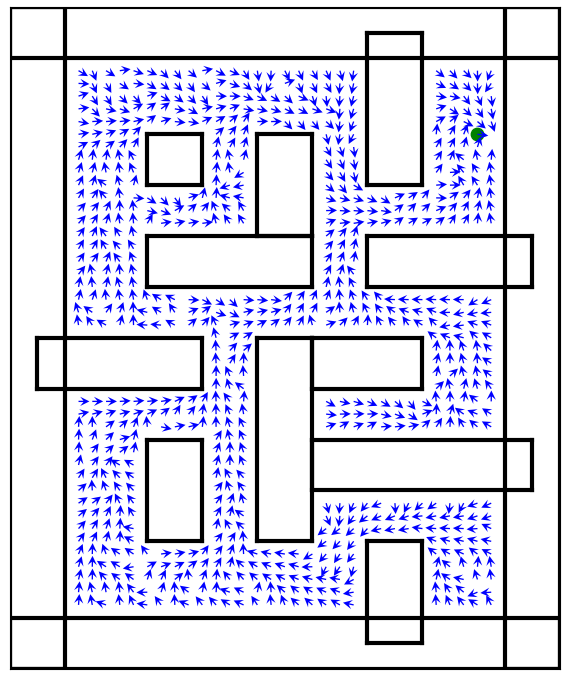}}
	\hfil
	\subfigure[\label{trajs_exp1}]{\includegraphics[width=1.1in]{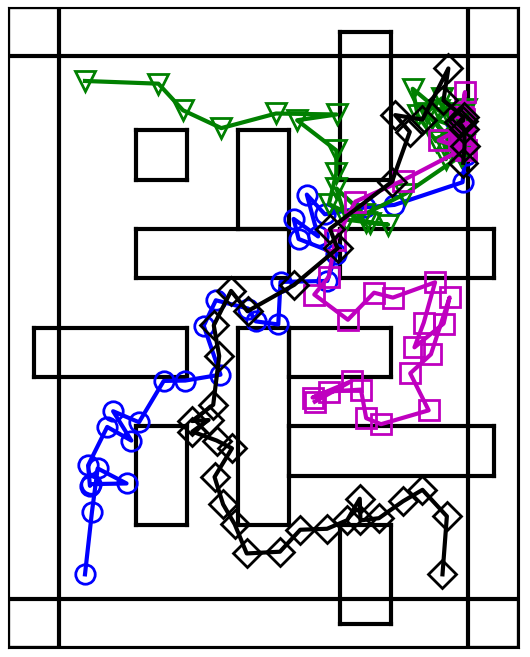}}
	\caption{Goal-fixed experiment. (a) The graph representation of place cells and synapses. (b) The activation function of the MSN population. (c) The gradient field of the MSN activation function. (d) Exemplary movement trajectories all converge to the goal location.}
	\label{experiment1b}
\end{figure*}

\textbf{Goal-fixed experiment.} Figure \ref{valMap_exp1} is the graph representation of place cells and their synapses. In the graph, each node represents a place cell with firing field center located at the position of the node. Each edge represents a pair of recurrent synapses with strength larger than a predefined threshold (e.g., 0.0001). The synaptic strengths among place cells well reproduces the complex layout of the maze. The color on each node characterizes the synaptic strength from this place cell to the population of MSN cells after replay. As our theory predicts,  the PC-MSN synaptic strength encodes the geodesic proximity between the firing field of each place cell and the firing field of the goal cell. Figure \ref{valFunc_exp1} shows the MSN activation as a function of the rat's location in the maze. The activation will ramp up if the rat is approaching the goal location from any initial locations, which conforms with experimental observations about MSN cells \cite{Meer2011, London2018}. Figure \ref{gradField_exp1} shows the gradient field of the MSN activation function. The gradient field has a clear tendency to approach the goal location. A rat that greedily chooses the gradient direction to move can navigate to the goal location from any initial locations. Figure \ref{trajs_exp1} shows several exemplary movement trajectories during test epochs.

\begin{figure*}[h]
	\centering
	\subfigure[\label{valMap_exp2}]{\includegraphics[width=1.35in]{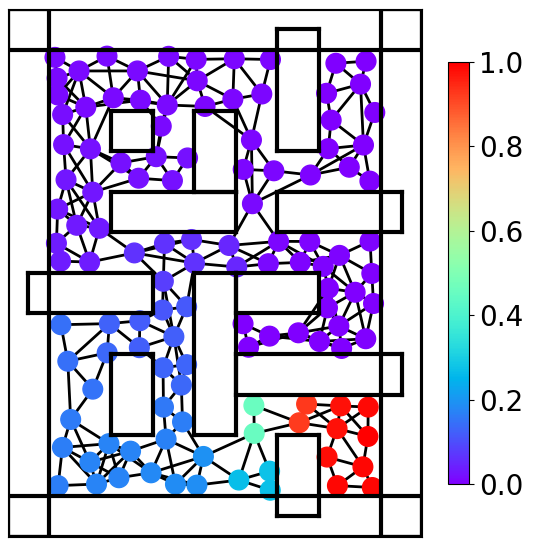}}
	\hfil
	\subfigure[\label{valFunc_exp2}]{\includegraphics[width=1.4in]{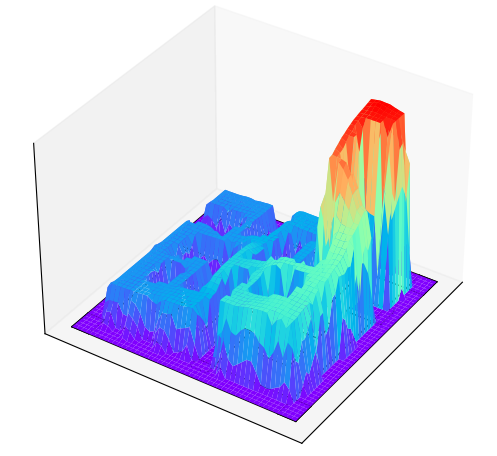}}
	\hfil
	\subfigure[\label{gradField_exp2}]{\includegraphics[width=1.1in]{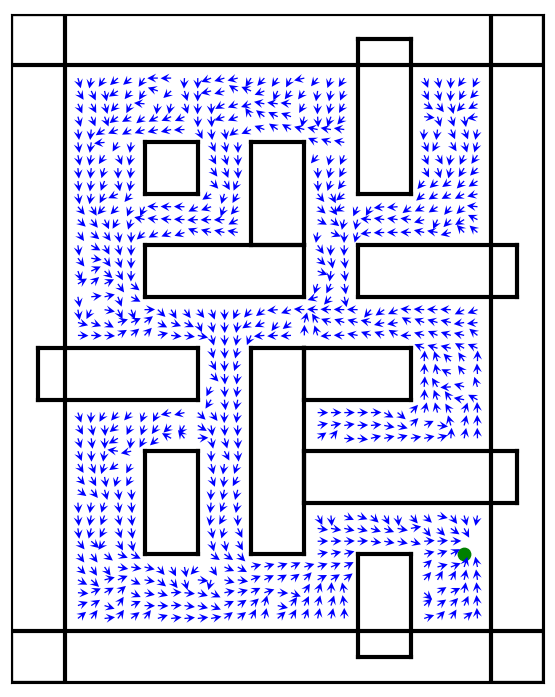}}
	\hfil
	\subfigure[\label{trajs_exp2}]{\includegraphics[width=1.1in]{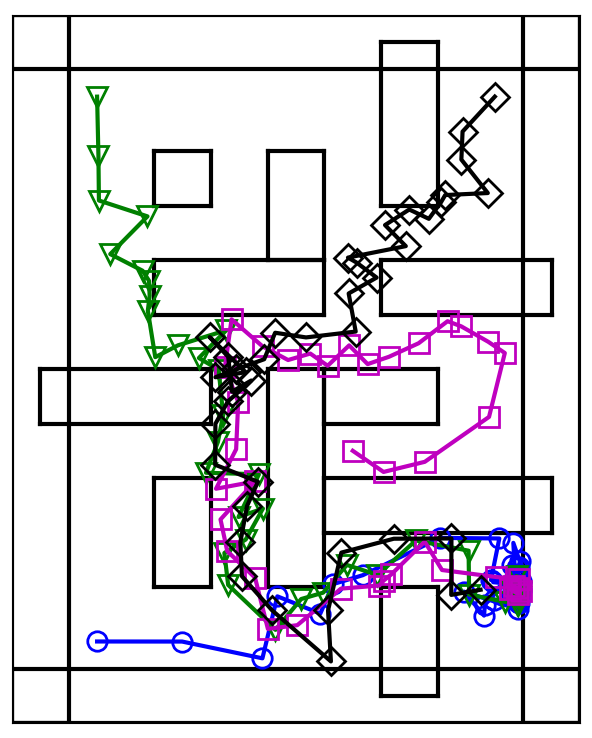}}
	\caption{Goal-changing experiment. (a) The graph representation after goal change. (b) The MSN activation function after goal change. (c) The gradient field of the MSN activation function. (d) Exemplary movement trajectories all converge to the new goal location.}
	\label{experiment2}
\end{figure*}

\textbf{Goal-changing experiment.} Figure \ref{valMap_exp2} shows the graph representation of place cells and their synapses after reperforming replay. The topology of the graph representation before goal changing is preserved after goal changing. However, the PC-MSN synaptic strengths are updated during replay to re-encode the geodesic proximity between each place cell to the new goal cell. As Figure \ref{valFunc_exp2} shows, the MSN activation function now ramps up along paths to the new goal location. Accordingly, the gradient field of the MSN activation function is re-directed toward the new goal location as shown in Figure \ref{gradField_exp2}. Figure \ref{trajs_exp2} shows several exemplary movement trajectories during test epochs. Starting from arbitrary initial locations, the rat can flexibly navigate to the new goal location without re-exploring the maze.

\begin{figure*}[h]
	\centering
	\subfigure[\label{pcfield_exp3}]{\includegraphics[width=1.05in]{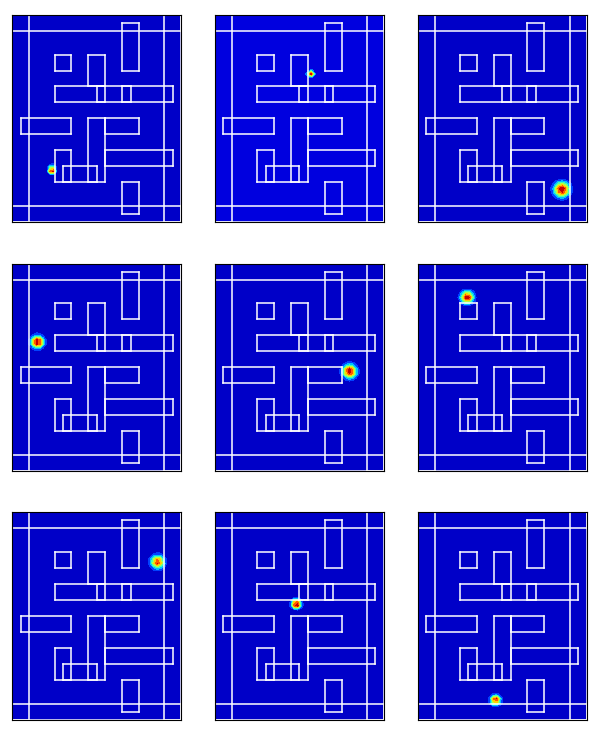}}
	\hfil
	\subfigure[\label{valMap_exp3}]{\includegraphics[width=1.3in]{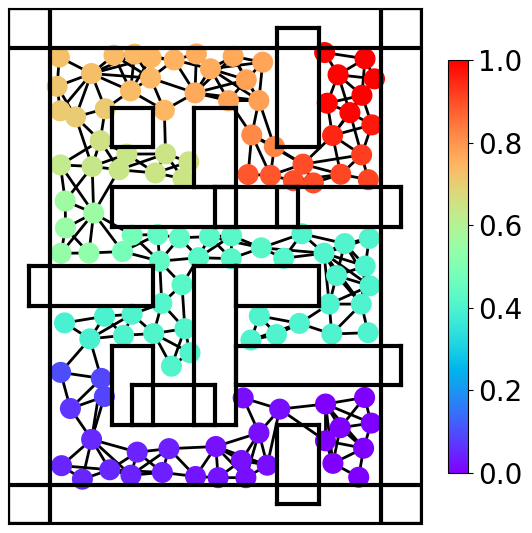}}
	\hfil
	\subfigure[\label{valFunc_exp3}]{\includegraphics[width=1.4in]{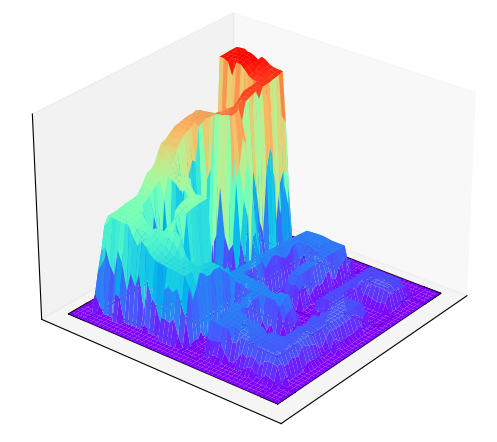}}
	\hfil
	\subfigure[\label{trajs_exp3}]{\includegraphics[width=1.1in]{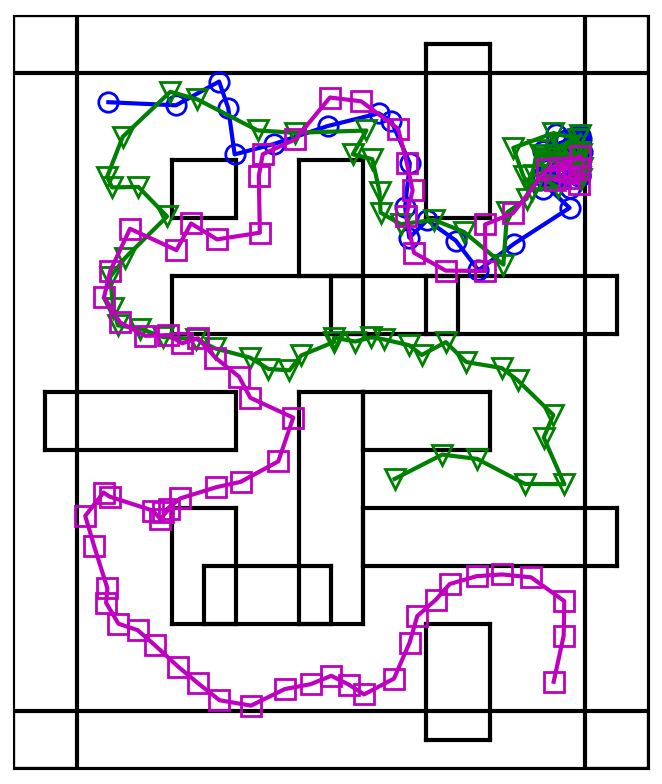}}
	\caption{Maze-changing experiment. (a) Exemplary place cell firing fields after re-exploration. (b) The graph representation after maze change. (c) The MSN activation function after maze change. (d) Exemplary movement trajectories show detour behavior.}
	\label{experiment3}
\end{figure*}

\textbf{Maze-changing experiment.} Figure \ref{pcfield_exp3} shows several updated place cell firing fields after re-exploration. Due to spatial layout change, place cells will no longer respond to the blocked areas. As Figure \ref{valMap_exp3} shows, the updated graph representation well characterizes the new spatial layout. Accordingly, the PC-MSN synaptic strengths reflect the changed geodesic proximity from each place cell to the goal cell. Similarly, the MSN activation function in Figure \ref{valFunc_exp3} ramps up along new paths to the goal location. 
Figure \ref{trajs_exp3} shows several exemplary movement trajectories in the new maze. Starting from initial positions far away from the goal location, the rat can find a detour to the goal location. Such detour behavior resembles that observed in mammals \cite{Alvernhe2011}.

\begin{figure*}[h]
	\centering
	\subfigure[\label{learnCurve_exp1}]{\includegraphics[width=1.8in]{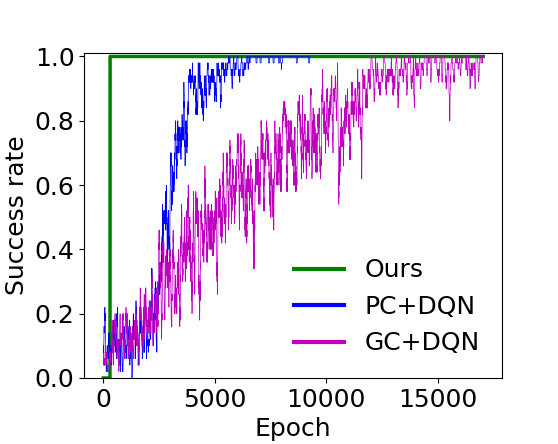}}
	\hfil
	\subfigure[\label{learnCurve_exp2}]{\includegraphics[width=1.8in]{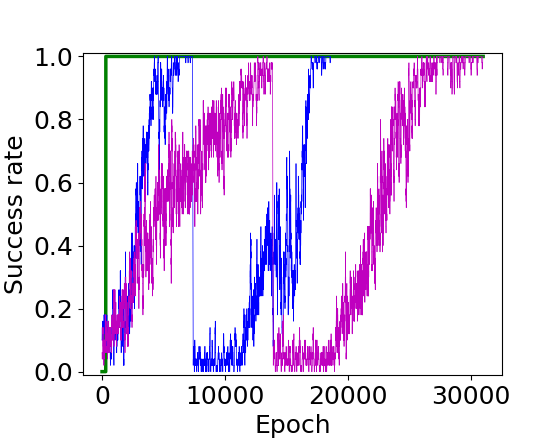}}
	\hfil
	\subfigure[\label{learnCurve_exp3}]{\includegraphics[width=1.8in]{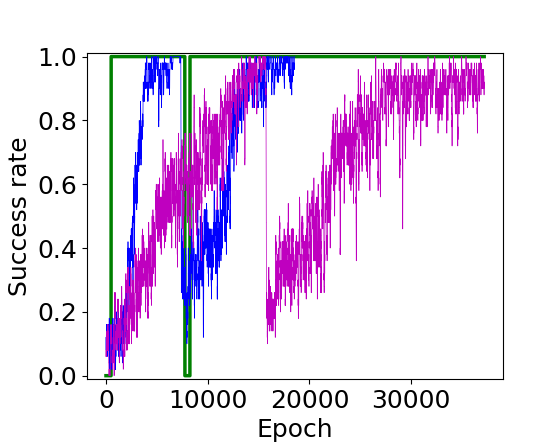}}
	\caption{Comparison to baselines. (a) Performance in the goal-fixed experiment. (b) Performance in the goal-changing experiment. (c) Performance in the maze-changing experiment.}
	\label{comparison}
\end{figure*}

\textbf{Comparison to baselines.} Figure \ref{learnCurve_exp1} shows the performance of baselines in the goal-fixed experiment. The success rate of the rat using our theory keeps at 1 after exploration epochs. In contrast, DQN-based rats require a large amount of epochs to incrementally improve performance by trial and errors. The GC+DQN shows slower progress than PC+DQN probably because it's more difficult to estimate the animal's position from grid cells' firing rates than from place cells' firing rates \cite{Wei2015}. 

Figure \ref{learnCurve_exp2} shows the performance of baselines in the goal-changing experiment. The success rate of the rat using our theory keeps at 1 despite the change of the goal location after 500 epochs. In contrast, the DQN-based rats suffer from a significant degradation of performance after goal changing at the 7750-th epoch and the 15500-th epoch for PC+DQN and GC+DQN, respectively. After goal changing, the trained state-policy mapping of DQN-based rats is unable to reach the new goal location. However, relearning a new mapping needs a large amount of trial and error epochs to gradually overwrite the learned mapping. 

Figure \ref{learnCurve_exp3} shows the performance of baselines in the maze-changing experiment. The success rate of the rat using our theory recovers to 1 after 100 re-exploration epochs at the 7750-th epoch. In contrast, the performance of DQN-based rats is significantly disrupted after the change of spatial layout at the 7750-th epoch and the 15500-th epoch for PC+DQN and GC+DQN, respectively. After spatial layout change, the trained state-policy mapping of DQN-based rats still tries to go through the blocked passages. However, relearning a correct mapping adapted to the new spatial layout still requires a large amount of trial and error epochs. Differently, the rat using our theory adapts to the new maze by updating the synaptic strengths among place cells to match the new spatial layout. With updated inter-PC synaptic strengths, PC-MSN synaptic strengths and the MSN activation function are immediately updated by replaying place cells. As a result, the rat can fast learn to reach arbitrary goal locations in the new maze by stochastically climbing the updated MSN activation function.
    \section{Discussion}
We have drawn inspiration from many neuroscience experiments and provided an integrative theory on how various information processing mechanisms in rodent brains support the efficient learning of flexible reward-seeking behavior. Such a systematic view about rodent brains is beneficial to the design of neuroscience-inspired artificial intelligence \cite{Hassabis2017}. We have shown that incorporating neuroscientific insights about learning, memory and motivation into the design loop of an artificial agent is important toward building artificial intelligence that matches the behavioral performance of animals.

Several works \cite{Sorscher2019, Cueva2018, Banino2018} trained recurrent neural networks (RNN) to perform path integration tasks in square environments. After the RNN correctly predicts positions from velocities, the firing fields of the hidden units of the RNN show grid-like patterns. The formation of grid patterns in these works typically requires some conditions, such as square environments without obstacles, place cells with firing fields given by the difference of Gaussians, special regularizations, etc. For a complex maze environment without these conditions satisfied, there is no guarantee that grid patterns will emerge. We leave the study of how grid-like patterns might emerge from RNN-based path integration in complex mazes for our future work. The work \cite{Banino2018} further fed the activations of RNN hidden units into another RNN for representing the policy and value function in actor-critic based RL. Similar to the DQN-based baselines,  learning reward-seeking behavior with the actor-critic RL is inefficient and the learned behavior is inflexible to changes of the rewarding location or the spatial layout.

The work \cite{Foster2000} proposed a linear actor-critic RL model and trained a simulated rat to perform Morris water maze tasks. One study \cite{Fremaux2013} used spiking neural networks to implement the actor-critic RL for Morris water-maze-like navigation tasks. It's a promising future direction to implement our theory with spiking neural networks. Another study \cite{Gustafson2011} found that using place cells with firing fields decaying with the geodesic distance from the field center can improve the efficiency of reinforcement learning. However, computing the geodesic firing fields of place cells requires complicated coordinate transformations of the original space. How such computations can be performed in brains remains unclear. Studies \cite{Stachenfeld2017, Stachenfeld2014} interpreted the firing field of a place cell as encoding the future expected number of visits to the firing field peak (center) from an initial location. Such predictive perspective on place cell firing fields can explain some asymmetric firing fields observed in previous experiments \cite{Mehta2000}. Our theory adopts the Gaussian model of place cell firing fields \cite{Foster2000, Gustafson2011, Banino2018, Sorscher2019, Muller1996}. It's an interesting direction to extend our theory based on other models of place cell firing fields.

Two studies \cite{Gerstner1997, Blum1996} proposed that the long-term potentiation (LTP) of synapses between place cells during exploration shifts the location encoded by the population activities away from the actual location of the animal. Such shifts are intepreted as a  vector field tending toward the goal location to guide the animal toward the goal. These studies can explain reward-seeking behavior solely based on activities of place cells. However, the exact neural representations of such displacement vectors (shifts) remain unclear. We show that it’s more biological plausible that the goal-directed vector field is encoded by the gradient field of the MSN activation. The work \cite{Muller1996} interpreted the hippocampus as a graph that connects place cells by synapses. This work used the biologically unrealistic Dijkstra's algorithm to compute the path with the smallest total synaptic resistance (reciprocal synaptic strength) between two place cells, which is further used to find the shortest path between two spatial positions. Compare to such path search algorithms, we show that the learning of PC-MSN synaptic strengths during hippocampal replay is a more biologically plausible model to explain how synaptic information stored among place cells is used to generate reward-seeking behavior.
    
    \bibliography{main}
    
    \appendix
    
    \section{Appendix}
\subsection{Derivations in Section 2.4}
\label{apdx1}
The chemical trace of the  $i$-th PC-MSN synapse at time $n\Delta t$ is given by,
\begin{equation}
	e^{(n\Delta t)}_i = \sum\limits_{k=0}^n \gamma^{(n-k) \Delta t} I_{c(k\Delta t)i}.
\end{equation}
Accordingly, the accumulated update to the $i$-th PC-MSN synapse when $n\to \infty$ can be written as follows, 
\begin{equation}
	\begin{aligned}
		\sum\limits_{n=0}^{\infty} \Delta V^{(n\Delta t)}_i &= \alpha \sum_{n=0}^{\infty} d^{(n\Delta t)} \sum_{k=0}^{n} \gamma^{(n-k)\Delta t} I_{c(k\Delta t)i}, \\
		&= \alpha \sum_{k=0}^{\infty} d^{(k\Delta t)} \sum_{n=0}^{k} \gamma^{(k-n)\Delta t} I_{c(n\Delta t)i}, \\
		&= \alpha \sum_{n=0}^{\infty} I_{c(n\Delta t)i} \sum_{k=n}^{\infty} \gamma^{(k-n)\Delta t} d^{(k\Delta t)}, \\
		&= \alpha \sum_{n=0}^{\infty} I_{c(n\Delta t)i} \sum_{k=n}^{\infty} \gamma^{(k-n)\Delta t} [g^{(k\Delta t)} + \gamma^{\Delta t} V^{(k\Delta t)}_{c[(k+1)\Delta t]} - V^{(k\Delta t)}_{c(k\Delta t)}], \\
		&= \alpha \sum_{n=0}^{\infty} I_{c(n\Delta t)i} (\sum_{k=n}^{\infty} \gamma^{(k-n)\Delta t} g^{(k\Delta t)} - V^{(n\Delta t)}_{c(n\Delta t)}), \\
		&= \alpha \sum_{n=0}^{\infty} I_{c(n\Delta t)i} (\sum_{k=n}^{\infty} \gamma^{(k-n)\Delta t} g^{(k\Delta t)} - V^{(n\Delta t)}_i).
	\end{aligned}
	\label{transfm}
\end{equation}
In Eq.~(\ref{transfm}), the second line is obtained by changing the role of $k$ and $n$ in the first line. The third line is obtained by interchanging the order of summation in the second line. The fourth line is obtained by plugging $d^{(k\Delta t)}$ in Eq.~(\ref{neoHeb}) into the third line. The fifth line is obtained by canceling out intermediate terms in the fourth line. The last line holds due to the indicator function $I_{c(n\Delta t)i}$. This derivation is inspired from the equivalence of forward and backward views of the TD($\lambda$) algorithm \cite{Sutton1998}.

The accumulated update in Eq.~(\ref{transfm}) equals the accumulated update of the following update rule,
\begin{equation}
	V^{[(n+1)\Delta t]}_i = V^{(n\Delta t)}_i + \alpha I_{c(n\Delta t)i} (\sum_{k=n}^{\infty} \gamma^{(k-n)\Delta t} g^{(k\Delta t)} - V^{(n\Delta t)}_i), \forall i.
\label{eqRule}
\end{equation}
Following Eq.~(\ref{eqRule}), the $i$-th PC-MSN synapse is updated only when the $i$-th place cell is activated. The change of synaptic strength is proportional to the difference between the future total discounted activation of the goal cell and the current synaptic strength. When $n$ is large, the synaptic strengths in Eq.~(\ref{neoHeb}) approximate $V^{(n\Delta t)}_i$ in Eq.~(\ref{eqRule}). Accordingly, the update rule in Eq.~(\ref{neoHeb}) is equivalent to a stochastic approximation process \cite{Nevelson1976} that solves the following equation as $n\to \infty$, 
\begin{equation}
	V_i^* = \mathbb{E}[\sum\limits_{n=0}^{\infty} \gamma^{n\Delta t}g^{(n\Delta t)}], \ \ c(0) = i,
	\label{goalGradA1}
\end{equation}
where the expectation $\mathbb{E}[\cdot]$ is taken with respect to all possible activation transitions among place cells during replay.

\subsection{Pretraining details}
\label{pretrain}
In order to train the policy that turns $\theta^\circ$ then runs forward, we use the following reward function,
\begin{equation}
	r = {vel}_g + 0.1 \vec{h}_g \cdot \vec{h}_r + 0.1 \vec{n}_z \cdot \vec{n}_r.
\end{equation}
Here ${vel}_g$ is the velocity of the rat along the targeted direction. $\vec{h}_g$ is the unit vector along the targeted direction. $\vec{h}_r$ is the unit vector along the head direction of the rat. $\vec{n}_z$ is the unit vector along the $z$-axis in the global coordinate frame. $\vec{n}_r$ is the unit vector along the $z$-axis in the local coordinate frame of the rat. To obtain high rewards, the rat requires to turn into the targeted direction and run as fast as possible while keeping $\vec{n}_r$ upward to avoid falling down. The TD3 algorithm trains each policy network from scratch to maximize the accumulated rewards collected from a behavior trajectory. Training each policy takes several hours on a GPU server.

\subsection{The model for computing grid cell firing rates}
\label{model-gc}
When the rat is located at position $x$, the firing rate of a grid cell is modeled by the sum of three 2-dimension cosine waves as follows \cite{Blair2007, Solstad2006},
\begin{equation}
	f_{gc}(x) = \sum_{k=1}^{3} \cos(\omega_k \cdot (x - p)),
\end{equation}
where $p$ is the spatial phase of the grid cell and $\omega_k$ is the wave vector of the $k$-th cosine wave. The directions of the wave vectors determine the angular orientation of the grid pattern. To produce a grid pattern with a given angular orientation $\theta$, the directions of the three vectors should take $\theta+\pi/2$, $\theta-\pi/6$ and $\theta+\pi/6$, respectively. The norms of the three wave vectors are the same and they determine the spatial periodicity of grid patterns. To produce a collection of grid cells \cite{Gustafson2011}, the values of $\theta$, $||\omega_k||$ and $p$ are taken from the set $[0, \pi/12, \pi/6, \pi/4]$, the set $[1, 2, 3, 4, 5]$ and the Cartesian product $[0, 0.5, 1]\times[0, 0.5, 1]$, respectively. 

\subsection{Hyperparameters}
\label{hyperparas}
	\begin{table}[h]
			\caption{Hyperparameters and their values}
			\label{para-table}
			\centering
			\begin{tabular}{ll}
					\toprule
					 Hyperparameter &  Value  \\
					\midrule
					Number of place cells & 120     \\
					Learning rate for Hebbian learning & 0.00001     \\
					Discount factor  & 0.999 \\
					Refractory period  & 0.01 \\
					Learning rate for TD learning  & 0.00001 \\
					Greediness level of behavior & 25 \\
					Number of sample directions & 20 \\
					Replay buffer size of DQN &  100000 \\
					Learning rate of DQN & 0.0003 \\
					Initial exploration coefficient of DQN & 0.6 \\
					Final exploration coefficient of DQN & 0.005 \\
					Discount factor of DQN & 0.99 \\
					Mini-batch size of DQN & 32 \\
					Number of layers of DQN & 5 \\
					Target net update rate of DQN & 0.01 \\
					Num of parallel workers of DQN & 32 \\
					Replay buffer size of TD3 &  100000 \\
					Learning rate of TD3 & 0.0001 \\
					Exploration coefficient of TD3 & 0.2 \\
					Discount factor of TD3 & 0.98 \\
					Mini-batch size of TD3 & 256 \\
					Number of layers of TD3 & 5 \\
					Target net update rate of TD3 & 0.01 \\
					Noise standard deviation of TD3 & 0.2 \\
					Noise clip coefficient of TD3 & 0.5 \\
					\bottomrule
				\end{tabular}
		\end{table}
	
\end{document}